\documentclass[reprint, prb,twocolumn,showkeys,showpacs,superscriptaddress]{revtex4-1}
\usepackage{braket}
\usepackage{amsmath,amssymb,times,color,graphicx,multirow,amsfonts,color}
\usepackage[colorlinks=true,citecolor=blue,linkcolor=blue]{hyperref}
\usepackage{physics,bm,here}
\usepackage{silence}
\begin{document}
\title{Difficulties in operator-based formulation of the bulk quadrupole moment}
\author{Seishiro Ono}
\affiliation{Institute for Solid State Physics, University of Tokyo, Kashiwa 277-8581, Japan}
\author{Luka Trifunovic}
\affiliation{RIKEN Center for Emergent Matter Science, Wako, Saitama 351-0198, Japan}
\author{Haruki Watanabe} 
\email{haruki.watanabe@ap.t.u-tokyo.ac.jp}
\affiliation{Department of Applied Physics, University of Tokyo, Tokyo 113-8656, Japan}

\begin{abstract}
Electric multipole moments are the most fundamental properties of
insulating materials.  However, the general formulation of bulk multipoles has
been a long standing problem. The solution for the electric dipole moment was
provided decades ago by King-Smith, Vanderbilt, and Resta. Recently, there have
been attempts at generalizing Resta's formula to higher-order multipoles. We
point out several issues in the recent proposals.
\end{abstract}
\maketitle

\section{Introduction}
Electric multipole moments are of fundamental importance in understanding
the property of insulators. Most significantly, they characterize the charge
distribution near boundaries of a finite sample. For two dimensional systems,
boundaries can be either one-dimensional edges or zero-dimensional corners.  A
nonzero electric polarization results in nonzero charge density for edges.
Recently, so-called quadrupole insulators that feature charged corners instead
of edges have attracted extensive research interest because of their relation
to higher-order topological
insulators.~\cite{parameswaran2017,schindler2018,peng2017,langbehn2017,song2017,Benalcazar61,PhysRevB.96.245115,fang2018,ezawa2018,shapourian2017,zhu2018,yan2018,wang2018,wang2018b,khalaf2018,khalaf2018b,trifunovic2019,nobuyuki2018}
To develop a systematic understanding of quadrupole insulators, we need to
establish a general framework that allows us to compute the quadrupole moment
of insulators.

Despite their importance, the precise formulation of multipole moments is known
to be a difficult task.  For finite systems under the open boundary condition (OBC), multipole moments can be simply defined by the classical
formula~\cite{Resta2007, Lines, kittel}:
\begin{align}
&n^{\text{(OBC)}}\equiv\frac{1}{V}\sum_{\bm x}\langle\Phi_0|\hat{n}_{\bm x}|\Phi_0\rangle,\label{eq:nOBC}\\
&p_{i}^{\text{(OBC)}}\equiv\frac{1}{V}\sum_{\bm x}x_i\langle\Phi_0|\hat{n}_{\bm x}|\Phi_0\rangle,\label{eq:pOBC}\\
&q_{ij}^{\text{(OBC)}}\equiv\frac{1}{V}\sum_{\bm x}x_ix_j\langle\Phi_0|\hat{n}_{\bm x}|\Phi_0\rangle.\label{eq:qOBC}
\end{align}
Here, $|\Phi_0\rangle$ is the many-body ground state, $\hat{n}_{\bm x}$ is the
number density operator at the position $\bm x$, and $V=L^2$ is the volume of
the system.  (Throughout this work, we consider a two dimensional square-shaped
system for brevity.) The polarization $p_{i}^{\text{(OBC)}}$ is independent
of the arbitrary choice of the origin only in neutral systems
($n^{\text{(OBC)}}=0$). Similarly, the quadrupole moment
$q_{ij}^{\text{(OBC)}}$ is well-defined only when all of $p_{i}^{\text{(OBC)}}$
($i=x,y$) and $n^{\text{(OBC)}}$ vanish.

The situation gets a lot more complicated when considering extended systems
under a periodic boundary condition (PBC), since the position operator in the
above expressions becomes ill-defined.~\cite{Martin,PhysRevLett.82.2560} For
band insulators, King-Smith and Vanderbilt formulated the bulk polarization in
terms of the Berry phase of Bloch
wavefunctions.\cite{PhysRevB.48.4442,PhysRevB.47.1651} The Berry phase approach
was generalized to many-body systems with interactions and/or disorders by
replacing the single-particle crystal momentum to the twisted angle of the
boundary condition.\cite{PhysRevB.49.14202,Souza,PhysRevX.8.021065}As an
alternative formulation, Resta~\cite{Resta} proposed the following
formula of the bulk electric polarization in many-body setting under a PBC:
\begin{align}
&p_i^{\text{(Resta)}}\equiv\frac{1}{2\pi}\text{Im}\ln\langle\Phi_0|\hat{U}_i|\Phi_0\rangle\mod1,\label{Resta}\\
&\hat{U}_i\equiv e^{2\pi i\hat{P}_i/L},\quad \hat{P}_i\equiv\sum_{\bm x}x_i\hat{n}_{\bm x}.\label{Ui}
\end{align}
According to Ref.~\onlinecite{Resta}, this relation holds even in the presence
of disorders and many-body interactions as long as the excitation gap is
non-vanishing. 

There have been several recent proposals on how to compute the bulk
quadrupole moment. For band insulators, the nested-Wilson loop
approach formulated in Refs.~\onlinecite{Benalcazar61,PhysRevB.96.245115} aims
at providing a way of computing the bulk contribution to the quantized corner
charge, under assumptions of the spatial symmetry and the so-called ``Wannier
gap''. We remark here that, although the nested-Wilson loop approach gives a
topological invariant, this invariant is generally not a \textit{bulk}
topological invariant---according to Ref.~\onlinecite{PhysRevB.96.245115} the
nested Wilson loop invariant can change its value if the Wannier gap is closed
while the bulk band gap and the protecting symmetry are maintained. Yet, the claim of
Refs.~\onlinecite{Benalcazar61,PhysRevB.96.245115} is that combining the
``bulk'' contribution to the edge polarization $p_{x}^{\text{edge}}$,
$p_{y}^{\text{edge}}$ obtained this way with an independent input on the corner
charge $Q^{\text{corner}}$ computed under a certain open boundary condition, one
gets the bulk quadrupole moment via
$q_{xy}=Q^{\text{corner}}-p_{x}^{\text{edge}}-p_{y}^{\text{edge}}$.~\cite{Benalcazar61,PhysRevB.96.245115}
As later pointed out by us,~\cite{trifunovic2019b} decorations by polarized
one-dimensional chains change the corner charge; the boundary Hamiltonian and thus
the boundary polarization are not generally defined quantities. Therefore it is not
possible to subtract the contribution from edge polarization.

As a more general definition of the bulk quadrupole moment in many-body
systems under a PBC, two independent
groups~\cite{Taylor,Ken} proposed a possible generalization of Resta's formula
to higher-order multipoles. For example, their formulas for quadrupole moments read
\begin{align}
&\tilde{q}_{ij}\equiv\frac{1}{2\pi}\text{Im}\ln\langle\Phi_0|\hat{U}_{ij}|\Phi_0\rangle\mod1,\label{qij}\\
&\hat{U}_{ij}\equiv e^{2\pi i\hat{Q}_{ij}/L^2},\quad \hat{Q}_{ij}\equiv\sum_{\bm x}x_ix_j\hat{n}_{\bm x}.\label{Uij}
\end{align}
Arguments supporting Eq.~\eqref{qij} are based on a field theoretical
calculation of the bulk response against a non-uniform electric field~\cite{Ken}
and a ``perturbation" theory~\footnote{We explain why this perturbation theory fails in Appendix~\ref{app1}.}  expanding the effect of the operator
$\hat{U}_{ij}$ in the series of $x_i x_j/L^2$.\cite{Taylor} The authors of
Refs.~\onlinecite{Ken,Taylor} have also provided a numerical proof of
Eq.~\eqref{qij} using tight-binding models of quadrupole insulators. More
recently, Refs.~\cite{agarwala2019,lin2019} employed the
formula~\eqref{qij} in their study of quadrupole insulators.

The goal of this work is to point out issues in the formula~\eqref{qij} for the bulk quadrupole moment.  A satisfactory formulation of the bulk quadrupole moment would fulfill the following requirements: (i) independence from the choice of origin and the period $L$ and (ii) quantization in the presence of sufficiently large point group symmetries so that it can serve as a topological invariant characterizing the quantized corner charge of quadrupole insulators.  However, we discuss that the operator $\hat{U}_{ij}$ in Eq.~\eqref{Uij} is inconsistent with the assumed PBC, and, as a consequence, $\tilde{q}_{ij}$ in
Eq.~\eqref{qij} meets none of these criteria.

\section{Problems in the proposed formula}\label{sec:quad}

\subsection{Violation of the periodicity}
\label{period}
We first point out an obvious issue in Eqs.~\eqref{qij} and \eqref{Uij}.  Let us consider a single-particle state $|\psi\rangle=\sum_{\bm x}\psi(\bm x)c_{\bm x}^\dagger|0\rangle$. The PBC requires that the wavefunction $\psi(\bm x)$ has the periodicity in $\bm{x}$ with
the period $L$ in both the $x$ and $y$ directions:
\begin{equation}
\psi(\bm x+L\bm{e}_i)=\psi(\bm x).\label{eq:period}
\end{equation}
where $\bm e_i$ represents the unit vector along $i=x,y$.  For example, the wavefunction of the Bloch state $\psi_{\bm{k}n}(\bm{x})=(1/\sqrt{V})u_{\bm
kn}(\bm{x})e^{i\bm k\cdot\bm{x}}$ fulfills this condition thanks to the quantization of $\bm{k}$ to the integer multiples of $2\pi/L$.

The operator $\hat{U}_i$ in Eq.~\eqref{Ui} preserves this periodicity. The
wavefunction of the state $\hat{U}_i|\psi\rangle$ is $e^{2\pi ix_i/L}\psi(\bm x)$, which satisfies Eq.~\eqref{eq:period}. In contrast, $\hat{U}_{ij}$ defined in Eq.~\eqref{Uij} violates the
periodicity.  The wavefunction of the state 
\begin{equation}
\hat{U}_{ij}|\psi\rangle=\sum_{\bm x}e^{2\pi i x_ix_j/L^2}\psi(\bm x)c_{\bm x}^\dagger|0\rangle
\end{equation}
is \emph{not} invariant under $\bm{x}\rightarrow\bm{x}+L\bm{e}_i$. Thus,
$\hat{U}_{ij}|\psi\rangle$ does not belong to the Hilbert space specified by the 
PBC.  Therefore the quantity
$\langle\psi|\hat{U}_{ij}|\psi\rangle$ (the inner-product of two
states $|\psi\rangle$ and $\hat{U}_{ij}|\psi\rangle$) lacks the physical
meaning. Reference~\onlinecite{Taylor} proposed to fix this issue by allowing for
discontinuities, but then the analyticity of the wavefunction would be lost.

To discuss more general states, let $\hat{T}_{L\bm{e}_i}$ be the translation operator that shifts $\bm{x}$ to $\bm{x}+L\bm{e}_i$ ($i=x,y$).
Imposing the PBC is equivalent to identifying $\hat{T}_{L\bm{e}_i}$ with an identity operator.
$\hat{U}_i$ in Eq.~\eqref{Ui} is consistent with
this identification because it commutes with $\hat{T}_{L\bm{e}_i}$. On the
other hand, $\hat{U}_{ij}$ in Eq.~\eqref{Uij} does not commute with
$\hat{T}_{L\bm{e}_i}$ and consequently it violates the boundary condition.

This simple discussion already poses a serious question about the physical meaning
of $\tilde{q}_{ij}$ in Eq.~\eqref{qij}.  Below we discuss the immediate consequence
of the lack of periodicity.

 \begin{figure}[t]
	\begin{center}
		\includegraphics[width=0.99\columnwidth]{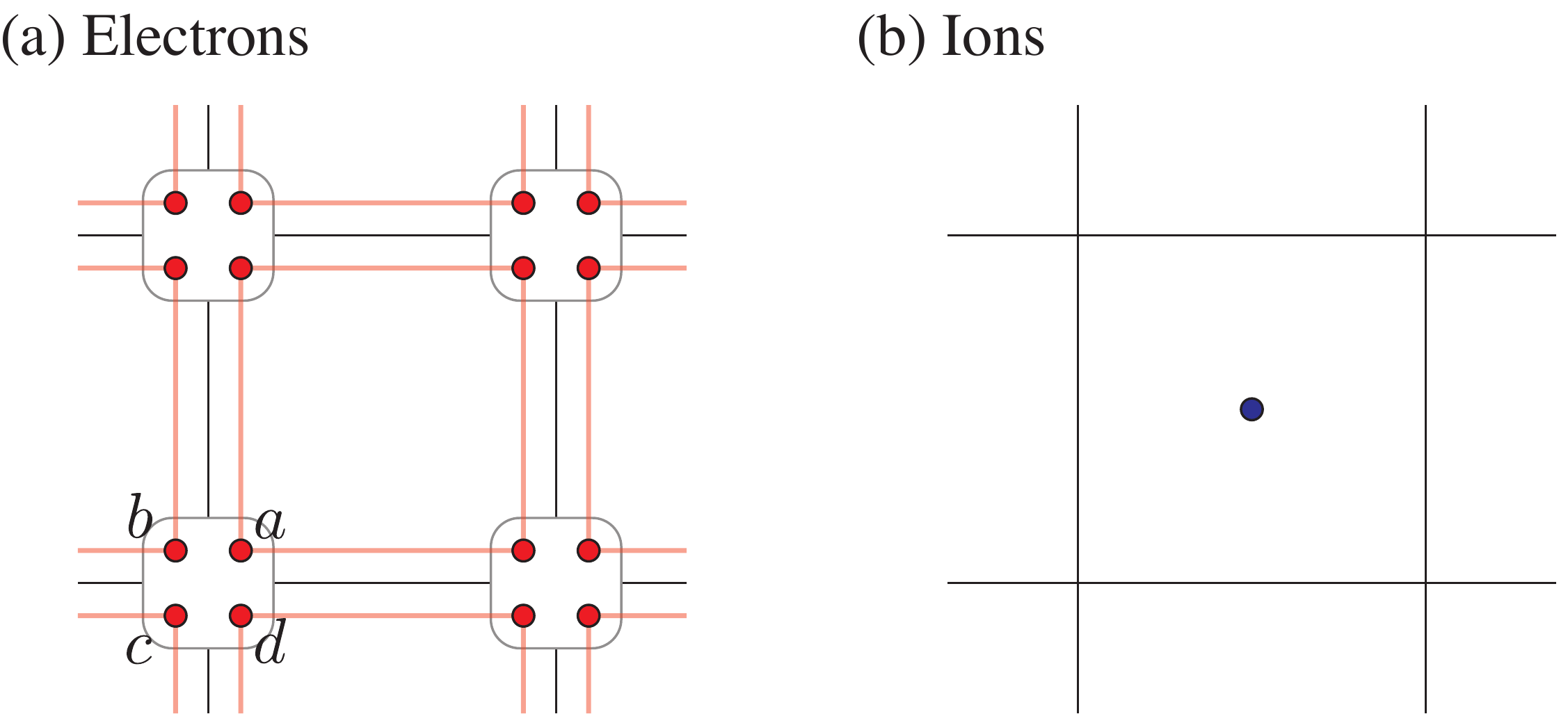}
		\caption{\label{fig} (a) The tight-binding
		model in Eq.~\eqref{eq:model2D1} describing the electronic contribution. Orange lines represent hopping. (b) Ions localized at plaquette center.
		}
	\end{center}
\end{figure}

\subsection{Tight-binding example of quadrupole insulator} 
\label{sec:2Dmodel}
Here we consider a simplified version of the four-band tight-binding model introduced in Ref.~\onlinecite{1809.02142}. 
See Appendix~\ref{moremodels} for more physical tight-binding models.
As illustrated in Fig.~\ref{fig} (a), the model has four orbitals ($a$, $b$, $c$, and $d$) at each lattice site.  The Hamiltonian in Fourier space reads
\begin{equation}
H_{\bm k}=-
\begin{pmatrix}
0&e^{ik_x}&0&e^{ik_y}\\
e^{-ik_x}&0&e^{ik_y}&0\\
0&e^{-ik_y}&0&e^{-ik_x}\\
e^{-ik_y}&0&e^{ik_x}&0\\
\end{pmatrix}.
\label{eq:model2D1}
\end{equation}
We assume $C_4$ rotation symmetry to see if $\tilde{q}_{ij}$ is quantized as proposed by Ref.~\onlinecite{Ken}.
\begin{equation}
U_{C_4}H_{k_x,k_y}=H_{-k_y,k_x}U_{C_4},\quad U_{C_4}\equiv \begin{pmatrix}
0&0&0&1\\
1&0&0&0\\
0&1&0&0\\
0&0&1&0\\
\end{pmatrix}.\label{uc4}
\end{equation}
We write the position of lattice sites as $\bm{R}=(x_n,y_m)$, where $x_n=x_1+n-1$ and $y_m=x_1+m-1$ ($n, m=1,2,\ldots,L$).  The lattice constant is set to be unity for simplicity. Here, $x_1$ is introduced to investigate the origin dependence of the results.~\footnote{See also Appendix of Ref.~\onlinecite{Ken} where the
dependence of $x_1$ is discussed using a tight-binding model.} Below we consider two familiar choices of $x_1$: $x_1=1$ ($x_L=L$) for every $L$ and $x_1=-(L-1)/2$ ($x_L=-x_1$) for an odd $L$.  

The four eigenvalues of $H_{\bm{k}}$ are $0$ (doubly degenerate) and $\pm
2$. The Bloch function of the lowest band ($\varepsilon_{\bm k}=-2$) reads $u_{\bm k}=(1,e^{-ik_x},e^{-i(k_x+k_y)},e^{-ik_y})^T/2$. 
In the Wannier basis, the ground state that
completely occupies the lowest band can be written $|\Phi_0\rangle=\prod_{\bm{R}}\hat{\gamma}_{\bm{R}}^\dagger|0\rangle$ with
\begin{equation}
\hat{\gamma}_{\bm{R}}^\dagger\equiv\frac{1}{2}(\hat{c}_{\bm
Ra}^\dagger+\hat{c}_{\bm{R}+\bm e_x,b}^\dagger+\hat{c}_{\bm{R}+\bm e_x+\bm
e_y,c}^\dagger+\hat{c}_{\bm{R}+\bm e_y,d}^\dagger).  \label{wannier}
\end{equation}
Using this real space expression, one can analytically evaluate $\tilde{q}_{ij}$.  We present the detailed calculation in Appendix~\ref{app2}.

Next we consider ionic contributions. To cancel the electric charge and polarization, we place an ion at every plaquette center, i.e. $\bm{x}=\bm{R}+(\bm{e}_x+\bm{e}_y)/2$ as illustrated in Fig.~\ref{fig} (b).  $\tilde{q}_{ij}$ in Eq.~\eqref{qij} for ions reduces to
\begin{equation}
\tilde{q}_{ij,\text{ions}}=\frac{1}{L^2}\sum_{\bm{R}}\left(R_i+\frac{1}{2}\right)\left(R_j+\frac{1}{2}\right).
\end{equation}
We subtract this reference value from the electronic contribution to impose the charge neutrality and vanishing polarization.
\begin{align}
&\Delta p_i^{\text{(Resta)}}=p_i^{\text{(Resta)}}-p_{i,\text{ions}}^{\text{(Resta)}}=0,\\
&\Delta \tilde{q}_{ij}=\tilde{q}_{ij}-\tilde{q}_{ij,\text{ions}}.
\end{align}

We summarize our results in Table~\ref{tab:model2D1}.  We immediately realize that the value of $\Delta\tilde{q}_{ij}$ depends sensitively on the detailed choice of origin [compare $x_1=1$ and $x_1=-(L-1)/2$ for an odd $L$] and also on the parity of $L$ [compare the even and odd $L$ for $x_1=1$]. Also, the value of $\Delta\tilde{q}_{xy}$ is not necessarily quantized [see the case of $x_1=1$ with an even $L$].

\begin{table}[t]
\caption{The analytic expression of $\Delta\tilde{q}_{ij}$ for the tight-binding model in Eq.~\eqref{eq:model2D1} in the limit of large $L$.  
\label{tab:model2D1}}
\begin{tabular}{c|ccc}
\hline\hline
$\,\,\,\,\,\,(x_1,L)\,\,\,\,\,\,$	& $\,\,\,\,\,\,(1,\text{even})\,\,\,\,\,\,$			&	$\,\,\,\,\,\,(1,\text{odd})\,\,\,\,\,\,$			&	$\,\,\,\,\,\,(-\frac{L-1}{2},\text{odd})\,\,\,\,\,\,$			\\\hline
$\Delta\tilde{q}_{xx}$, $\Delta\tilde{q}_{yy}$	&	$\frac{1}{4}$	&	$\frac{1}{4}$	&	$\frac{3}{4}$\\
$\Delta\tilde{q}_{xy}$, $\Delta\tilde{q}_{yx}$	&	$\frac{1}{2}+\frac{1}{\pi}\arctan\frac{2}{\pi}\footnote{$=0.680454\cdots$}$	&	$\frac{1}{2}$	&	$\frac{3}{4}$\\
\hline\hline
\end{tabular}
\end{table}

\section{Difficulties in the improvement}
\subsection{Absence of Resta's type formula}
Let us ask if one can fix the issues in $\tilde{q}_{ij}$ identified above.
Here we explore the Resta-type expression $\tilde{q}'\equiv\frac{1}{2\pi}\text{Im}\ln\langle\Phi_0|\hat{U}'|\Phi_0\rangle$ with $\hat{U}'\equiv e^{2\pi i\sum_{\vec{x}}\theta_{\vec{x}}\hat{n}_{\vec{x}}}$.
To be consistent with the PBC, $\theta_{\vec{x}}$ must have the following form
\begin{equation}
\theta_{\bm{x}}=\frac{m_xx+m_yy}{L}+\vartheta_{\bm{x}}\quad (m_i\in\mathbb{Z}),
\end{equation}
where $\vartheta_{\bm{x}}\in\mathbb{R}$ is a periodic function of $\bm{x}$. 
In particular, quadratic terms such as $x_ix_j/L^2$ cannot appear in $\theta_{\bm{x}}$ contrary to Eq.~\eqref{Uij}.

One may introduce a cut-off function $C_{\vec{x}}\in[0,1]$ that is $1$ (constant) when $\bm{x}$ is far away from the boundary (i.e., $x_1\ll x\ll x_L$ and $y_1\ll y\ll y_L$) and smoothly approaches to $0$ when $\vec{x}$ is near the boundary with an intermediate length scale $\ell$ ($1\ll \ell \ll L$).  For example, we can use
\begin{eqnarray}
C_{\vec{x}}=&&\tfrac{1+\tanh\frac{x-x_1-\frac{L}{4}}{\ell}}{2}\tfrac{1-\tanh\frac{x-x_L+\frac{L}{4}}{\ell}}{2}\notag\\
&&\times\tfrac{1+\tanh\frac{y-y_1-\frac{L}{4}}{\ell}}{2}\tfrac{1-\tanh\frac{y-y_L+\frac{L}{4}}{\ell}}{2}.
\end{eqnarray}
Then $\vartheta_{\bm{x}}= C_{\bm{x}}\,xy/L^2$ becomes effectively a periodic function when terms proportional to $e^{-L/4\ell}$ are neglected. However, we found that such a modification does not resolve the issues, because the contribution to $\tilde{q}'$ from the decaying region ($|\partial_xC_{\bm{x}}|\sim 1/\ell$) is non-neglegible and spoils the bulk contribution from the interior ($C_{\bm{x}}\simeq 1$).

\subsection{Absence of a Berry-phase type formula}
Next, let us examine Berry-phase type formulas.  According to the modern theory of electric polarization,\cite{PhysRevB.48.4442,PhysRevB.47.1651} the Berry phase 
\begin{equation}
\int_{\text{BZ}}\frac{d^2k}{(2\pi)^2}\sum_{n=1}^{N_{\text{occ}}}i\langle u_{\bm{k}n}| \partial_{k_i}u_{\bm{k}n}\rangle\label{KSV1}
\end{equation}
gives the electric polarization. Here, $|u_{\bm{k}n}\rangle$ is the Bloch function of $n$-th occupied band.  This Berry phase can be interpreted as the expectation value of the position operator measured from the origin of the unit cell $\int d^2x\sum_{n=1}^{N_{\text{occ}}} w_{n\bm{R}}(\bm{x})^* (x_i-R_i) w_{n\bm{R}}(\bm{x})$ of the Wannier state $w_{n\bm{R}}(\bm{x})\equiv L^{-2}\sum_{\bm{k}}e^{i\bm{k}\cdot(\bm{x}-\bm{R})}u_{\bm{k}n}(\bm{x})$.  
Therefore one may guess that the bulk quadrupole is given by $\int d^2x\sum_{n=1}^{N_{\text{occ}}}w_{\bm{R}}(\bm{x})^* (x_i-R_i)(x_j-R_j)w_{\bm{R}}(\bm{x})$, i.e.,
\begin{equation}
\int_{\text{BZ}}\frac{d^2k}{(2\pi)^2}\sum_{n=1}^{N_{\text{occ}}}\langle\partial_{k_i}u_{\bm{k}n} |\partial_{k_j}u_{\bm{k}n}\rangle.\label{KSV2}
\end{equation}
However, this cannot be the case because the quantity in Eq.~\eqref{KSV2} is not invariant under the gauge transformation $w_{\vec{k}}\in\text{U}(N_{\text{occ}})$ among occupied bands,
\begin{equation}
|u_{\bm{k}n'}\rangle'=\sum_{n=1}^{N_{\text{occ}}} |u_{\bm{k}n}\rangle(w_{\bm{k}})_{n,n'}.\label{unitary}
\end{equation}
We can fix the gauge-invariance by inserting the projector onto unoccupied bands $1-P_{\bm{k}}$~\cite{Marzari,Resta2,Souza}
\begin{equation}
T_{ij}\equiv\int_{\text{BZ}}\frac{d^2k}{(2\pi)^2}\sum_{n=1}^{N_{\text{occ}}}\langle \partial_{k_i}u_{\bm{k}n}| (1-P_{\bm{k}})|\partial_{k_j}u_{\bm{k}n}\rangle.
\end{equation}
However, we then found that this quantity does not produce a useful topological invariant, because $T_{ij}$ is always proportional to an identity matrix in the presence of $n=3$, $4$, or $6$-fold rotation symmetry. [This is because it satisfies $\sum_{i,j}(p)_{i'i}T_{ij}(p^T)_{jj'}=T_{i'j'}$ under a point group symmetry $p\in \text{O}(2)$.] A useful topological invariant would be constructed from an integral with an integer ambiguity, but, to our knowledge, the Berry phase in Eq.~\eqref{KSV1} is the only combination with that property in two dimensions.

\section{Summary}
We have analyzed the definition of the bulk quadrupole moment independently
proposed by two groups.~\cite{Ken,Taylor} We find that the proposed definition
of the bulk quadrupole moment fails even for a simple non-interacting example.
Our analysis reveals that the issues with the proposed definition are related
to violation of periodicity. Possible strategies to fix these issues all seem
to fall short.

The obstacles in obtaining a generalization of Resta's formulation of bulk
polarization to higher multipoles are perhaps best illustrated by our findings
that even a simpler task, namely, finding a formulation for single-particle
systems seems to fail. Although we cannot provide a general proof that such
single-particle formulation of bulk quadrupole moment does not exist, we
give a strong indication that this task may be a difficult one.

A proper definition of higher multiples in crystals and formulas that allow
practical calculations are topics of broad interest, not limited to
computational and theoretical solid state physics. Despite the fact that our
findings support in some sense a `no-go' statement, we hope that this work
will serve as the first step toward the future resolution to defining bulk
multiple moments.

\begin{acknowledgements}
The authors would like to thank A.
Furusaki, M. Oshikawa, and A. Shitade for useful discussions.  We would also
like to acknowledge the communications with authors of
Refs.~\onlinecite{Ken,Taylor} prior to the submission of this manuscript. The
work of S.O. is supported by the Materials Education program for the future leaders
in Research, Industry, and Technology (MERIT).  The work of H. W. is supported
by JSPS KAKENHI Grant No.~JP17K17678 and by JST PRESTO Grant No.~JPMJPR18LA. 
\end{acknowledgements}

\appendix

\section{Subtlety in the perturbative expansion}
\label{app1}
Resta~\cite{Resta} used the ``first-order perturbation theory'' to derive Eq.~\eqref{Resta} for interacting systems. Ref.~\onlinecite{Taylor} used a similar perturbative argument to verify the formula~(\ref{Uij}). Here, we review an issue with such perturbative treatment following Ref.~\onlinecite{PhysRevX.8.021065}.  Only in this appendix do we assume a general $d$ dimensional system with the period $L_i$ in $i$-th direction.  According to Resta,~\cite{Resta} 
\begin{equation}
\hat{U}_x|\Phi_0\rangle\doteq e^{i\gamma}\left(|\Phi_0\rangle+\frac{2\pi}{L_x}\sum_{N>0}|\Phi_N\rangle\frac{\langle\Phi_N|\hat{J}_x|\Phi_0\rangle}{E_{N}-E_0}+\cdots\right),\label{first}
\end{equation}
where $\hat{J}_x$ is the sum of the $x$-component of the current operators over
the entire space. If the above relation were a controlled expansion in the
series of $L_x^{-1}$, the expectation value
$\langle\Phi_0|\hat{U}_x|\Phi_0\rangle$ would behave as
\begin{equation}
|\langle\Phi_0|\hat{U}_x|\Phi_0\rangle|\doteq 1+O(L_x^{-2}),\label{wrong}
\end{equation}
which converges to $1$ in the large $L_x$ limit.
However, this turns out not to be the case because of the volume sum hidden in
$\hat{J}_x$.~\cite{PhysRevX.8.021065} We put the dot over the equality in
Eqs.~\eqref{first} and ~\eqref{wrong} as a caution to the reader.

This issue can be readily seen in the case of band
insulators. The matrix element of $\hat{U}_x$ among Bloch states is given by 
\begin{equation}
(\hat{U}_x)_{\bm k'n',\bm k n}=\delta_{\bm k',\bm k+(2\pi/L_x)\bm e_x}(B_{\bm k})_{n'n},\label{matrixelem}
\end{equation}
where $n=1,2,\cdots,N_{\text{occ}}$ is the band index of occupied bands and $B_{\bm k}$ is an $N_{\text{occ}}$-dimensional matrix representing the discretized Berry connection defined by
\begin{align}
(B_{\bm k})_{n'n}&\equiv \langle u_{\bm k+(2\pi/L_x)\bm e_xn'}|u_{\bm k n}\rangle.
\end{align}
By expanding $B_{\bm k}$ to the second order in $L_x^{-1}$, we find
\begin{equation}
|\langle\Phi_0|\hat{U}_x|\Phi_0\rangle|=e^{-2\pi^2 G_{xx} V/L_x^2+O(V/L_x^3)},\label{correctscale}
\end{equation}
where $V=L_xL_y\cdots$ is the volume and $G^{ij}$ is the quantum metric tensor defined by~\cite{Marzari,Resta2,Souza}
\begin{align}
&G_{ij}\equiv \text{Re} \int\frac{d^dk}{(2\pi)^d}(g_{\bm{k}})_{ij},\label{QM}\\
&(g_{\bm{k}})_{ij}=\sum_{n=1}^{N_{\text{occ}}}\langle\partial_{k_i} u_{\bm kn}|(1-P_{\bm{k}}) |\partial_{k_j} u_{\bm kn}\rangle.\label{gij2}
\end{align}
Note that $G_{ij}=O(1)$, i.e., it does not depend on the system
size.  For example, for the isotropic case in which all
$L_i$'s are identical to $L$, the right-hand side of Eq.~\eqref{correctscale}
is $e^{-2\pi^2 G^{xx}L^{{d-2}}+O(L^{{d-3}})}$.  In the large $L$ limit, it converges
to a number in the range $0$ and $1$ in two dimension, and it vanishes in higher dimensions.  This is in sharp contrast to the
behavior in Eq.~\eqref{wrong}.  Thus the first-order perturbation~(\ref{first})
does not hold in general.~\footnote{Even when $d=1$, Eq.~\eqref{wrong} is still
violated because the exponent of Eq.~\eqref{correctscale} is $O(L_x^{-1})$, not
$O(L_x^{-2})$.} 

Quite remarkably, despite the lack of a general proof of Eq.~\eqref{Resta} for
interacting systems in multi-dimensions, to best of our knowledge there is no
counterexample to Eq.~\eqref{Resta}.  Unfortunately, as we discuss in the main text, the
circumstances are not so favorable for the validity of the proposed
expression~(\ref{Uij}), where a similar perturbative expansion argument was
used.~\cite{Taylor}

\section{Details on the tight-biding calculation}
\label{app2}
Here we present the derivation of the result for the model in Eq.~\eqref{eq:model2D1} summarized in Table~\ref{tab:model2D1}.  To compute $\tilde{q}_{ij}$ in Eq.~\eqref{qij}, one has to compute the matrix element of $\hat{U}_{ij}$ in Eq.~\eqref{Uij} among occupied single-particle states. 
To this end, working in the Wannier basis instead of the Bloch basis is advantageous, because all the off-diagonal elements [i.e. $(\hat{U}_{ij})_{\bm{R},\bm{R}'}$, $\bm{R}\neq\bm{R}'$] vanish for the Wannier state in Eq.~\eqref{wannier}. Because of this nice property, we have
\begin{equation}
\tilde{q}_{ij}=\frac{1}{2\pi}\sum_{\bm{R}}\text{Im}\ln(\hat{U}_{ij})_{\bm{R},\bm{R}}.\label{simplified}
\end{equation}
The expression for the diagonal matrix element $(\hat{U}_{ij})_{\bm{R},\bm{R}}$ depends on the position of plaquettes, i.e.,
whether $\bm{R}$ is in the bulk or around the boundary. Introducing the shorthand
notation $F_z=e^{2\pi i z/L^2}$, the diagonal elements can be written as
\begin{align}
	&(\hat{U}_{xx})_{\bm{R},\bm{R}}=\frac{1}{2}\times
\begin{cases}
	F_{R_x^2}+F_{(R_x+1)^2}&\text{for } R_x<x_L\\
	F_{x_L^2}+F_{x_1^2}&\text{for } R_x=x_L
\end{cases},\label{Mxxc}\\
	&(\hat{U}_{yy})_{\bm{R},\bm{R}}=\frac{1}{2}\times
\begin{cases}
	F_{R_y^2}+F_{(R_y+1)^2}&\text{for } R_y<y_L\\
	F_{y_L^2}+F_{y_1^2}&\text{for } R_y=y_L
\end{cases},\label{Myyc}
\end{align}
and
\begin{widetext}
\begin{equation}
	(\hat{U}_{xy})_{\bm{R},\bm{R}}=(\hat{U}_{yx})_{\bm{R},\bm{R}}=\frac{1}{4}\times
\begin{cases}
	F_{R_xR_y}+F_{(R_x+1)R_y}+F_{(R_x+1)(R_y+1)}+F_{R_x(R_y+1)}&\text{for } R_x<x_L\text{ and }R_y<y_L\\
	F_{x_LR_y}+F_{x_1R_y}+F_{x_1(R_y+1)}+F_{x_L(R_y+1)}&\text{for } R_x=x_L\text{ and }R_y<y_L\\
	F_{R_xy_L}+F_{(R_x+1)y_L}+F_{(R_x+1)y_1}+F_{R_xy_1}&\text{for } R_x<x_L\text{ and }R_y=y_L\\
	F_{x_Ly_L}+F_{x_1y_L}+F_{x_1y_1}+F_{x_Ly_1}&\text{for } R_x=x_L\text{ and }R_y=y_L
\end{cases}.\label{Mxyc}
\end{equation}
To reproduce Table~\ref{tab:model2D1}, one just has to plug these expressions into Eq.~\eqref{simplified} and extract the value in the large $L$ limit.
\end{widetext}

 \begin{figure}
	\begin{center}
		\includegraphics[width=0.99\columnwidth]{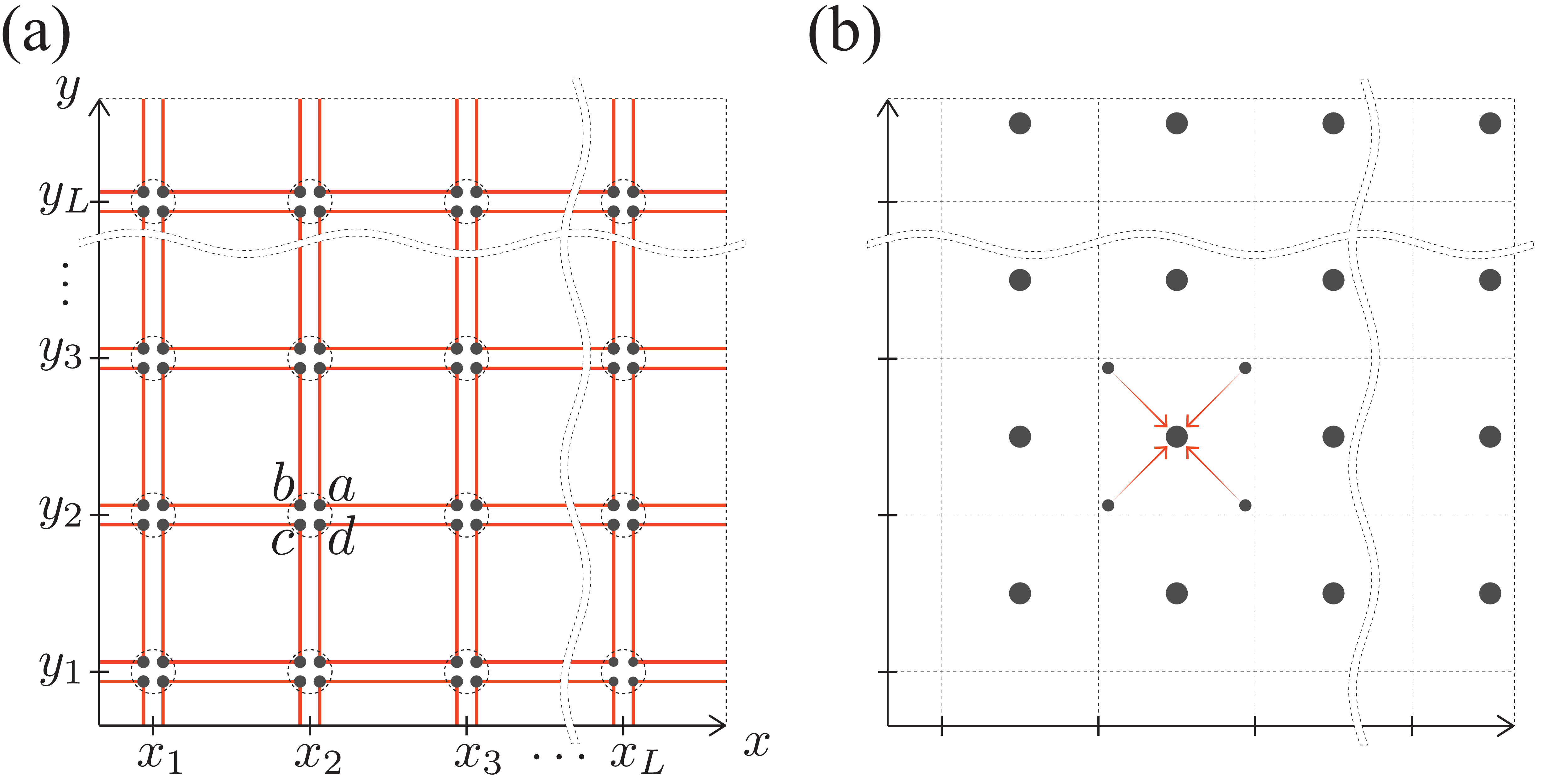}
		\caption{\label{m1} Illustration of the first model (a) and the atomic limit (b).}
	\end{center}
\end{figure}

\section{Examples of tight-biding models}
\label{moremodels}
In this section, we discuss two tight-binding models that are more physically natural than the simplified one discussed in Sec.~\ref{sec:2Dmodel}. This exercise will support our claim, clarifying the issues in the proposed formula~\eqref{qij}.

\subsection{Model 1}
\label{app:m1}
The first model is illustrated in Fig.~\ref{m1} (a). As far as the electronic degrees of freedom are concerned, this model is identical to the one in Eq.~\eqref{eq:model2D1}.  The only difference is the position of ions. Since lattice sites are located at $\bm{R}=(x_n,y_m)$ (see the main text), here we assume that one ion sits at every lattice site $\bm{R}$. This choice gives the reference value
\begin{align}
\tilde{q}_{xy,\text{ions}}&=\frac{(2x_1+L-1)^2}{4}.
\end{align}

The lowest band of the model \eqref{eq:model2D1} has a Wannier orbital centering at $\bm{x}=\bm{R}+(\frac{1}{2},\frac{1}{2})$ for each $\bm{R}$.
The band insulator that completely occupies this band has
\begin{equation}
\tilde{q}_{xy}^{\text{(a)}}=\frac{1}{2\pi}\sum_{\bm{R}}\text{Im}\ln(\hat{U}_{xy}^{\text{(a)}})_{\bm{R},\bm{R}},
\end{equation}
where $(\hat{U}_{xy}^{\text{(a)}})_{\bm{R},\bm{R}}$ is given in Eq.~\eqref{Mxyc}. We tabulate the value of $\Delta\tilde{q}_{xy}^{\text{(a)}}=\tilde{q}_{xy}^{\text{(a)}}-\tilde{q}_{xy,\text{ions}}$ in Table~\ref{tab:m1}.

We compare this to the atomic limit illustrated in Fig.~\ref{m1} (b). In this limit, all electrons are strictly localized at the Wannier center $\bm{x}=\bm{R}+(\frac{1}{2},\frac{1}{2})$ for every $\bm{R}$. The above band insulator can be adiabatically connected to this limit without closing the bulk gap or breaking the $C_4$ symmetry.  For the atomic limit, we find
\begin{align}
&(\hat{U}_{xy}^{\text{(b)}})_{\bm{R},\bm{R}}=e^{2\pi i(R_x+1/2)(R_y+1/2)/L^2},\label{Mxyd}\\
&\tilde{q}_{xy}^{\text{(b)}}=\frac{1}{2\pi}\sum_{\bm{R}}\text{Im}\ln(\hat{U}_{xy}^{\text{(b)}})_{\bm{R},\bm{R}}=\frac{(2x_1+L)^2}{4}.
\end{align}
We tabulate the value of $\Delta\tilde{q}_{xy}^{\text{(b)}}=\tilde{q}_{xy}^{\text{(b)}}-\tilde{q}_{xy,\text{ions}}$ in Table~\ref{tab:m1}.  Clearly $\Delta\tilde{q}_{xy}^{\text{(b)}}$ does not agree with the atomic limit $\Delta\tilde{q}_{xy}^{\text{(a)}}$, despite the fact that they are smoothly connected to each other. This implies that $\Delta\tilde{q}_{xy}$ cannot, in general, serve as the topological invariant.

\begin{table}
\caption{Analytic expression of $\Delta\tilde{q}_{xy}$ for the first model in the limit of large $L$.
\label{tab:m1}}
\begin{tabular}{c|ccc}
\hline\hline
$(x_1,L)$	& $(1,\text{even})$			&	\quad$(1,\text{odd})$\quad\quad &	\quad$(-\frac{L-1}{2},\text{odd})$\quad\quad			\\\hline
 $\Delta\tilde{q}_{xy}^{\text{(a)}}$	&	$\frac{1}{4}+\frac{1}{\pi}\arctan\frac{2}{\pi}\footnote{$=0.430454\cdots$}$	&	$\frac{3}{4}$	&	$0$\\
 $\Delta\tilde{q}_{xy}^{\text{(b)}}$	&	$\frac{3}{4}$	&	$\frac{1}{4}$	&	$\frac{1}{4}$\\
\hline\hline
\end{tabular}
\end{table}

 \begin{figure}
	\begin{center}
		\includegraphics[width=0.99\columnwidth]{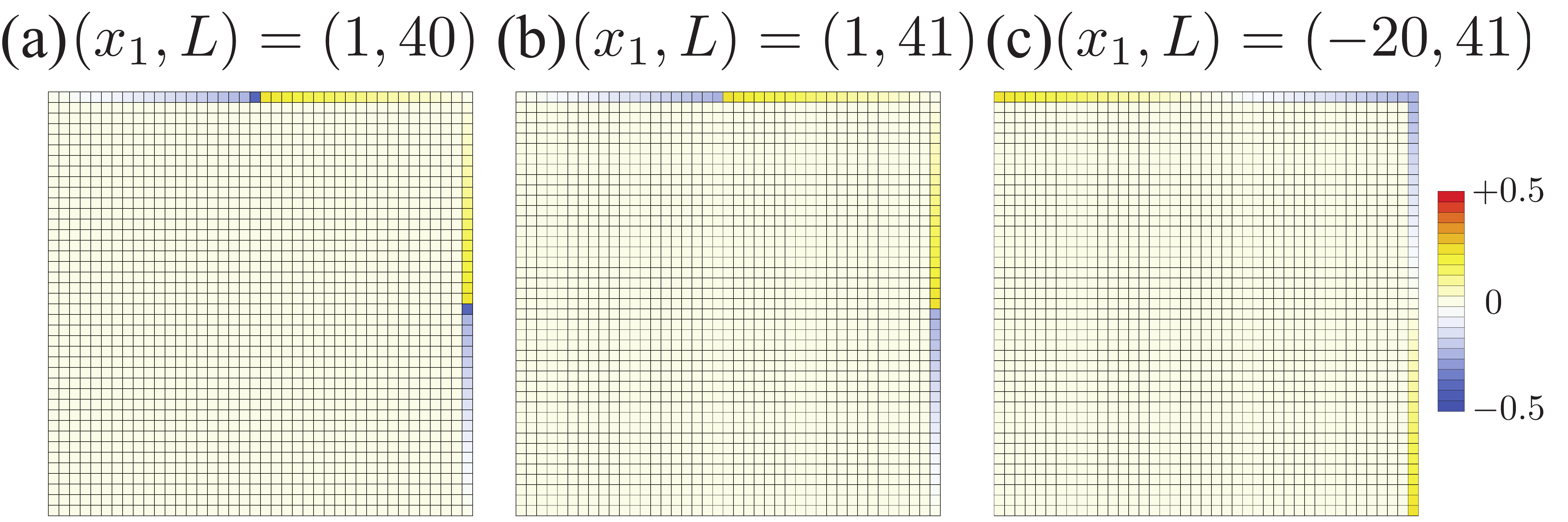}
		\caption{\label{m1s} The plot of $\Delta_{\bm{R}}$ in
		Eq.~\eqref{deltaR} for the three settings of $(x_1,L)$. In the panel (a), the two plaquettes at
		$\bm{R}=(L,L/2)$ and $(L/2,L)$ produce the irrational
		contribution in Table~\ref{tab:m1}.}
	\end{center}
\end{figure}

To pin down the origin of the mismatch, we compare $(\hat{U}_{xy}^{\text{(a)}})_{\bm{R},\bm{R}}$ and $(\hat{U}_{xy}^{\text{(b)}})_{\bm{R},\bm{R}}$ by plotting
\begin{equation}
\Delta_{\bm{R}}\equiv\frac{1}{2\pi}\text{Im}\ln\frac{(\hat{U}_{xy}^{\text{(a)}})_{\bm{R},\bm{R}}}{(\hat{U}_{xy}^{\text{(b)}})_{\bm{R},\bm{R}}}\mod 1\label{deltaR}
\end{equation}
as a function of $\bm{R}$.  Figure~\ref{m1s} shows 
that the discrepancy originates purely from the boundary. In fact, the sum of
$\Delta_{\bm{R}}$ over the boundary (i.e. $R_x=x_L$ or $R_y=y_L$) precisely
accounts for the difference
$\Delta\tilde{q}_{xy}^{\text{(a)}}-\Delta\tilde{q}_{xy}^{\text{(b)}}$.
Furthermore, $(\hat{U}_{xy})_{\bm{R},\bm{R}}^{\text{(b)}}$ for the plaquettes
near the boundary has a small amplitude and $|\langle
\Phi_0|\hat{U}_{xy}|\Phi_0\rangle|$ decays exponentially with the system size
($e^{-c L}$ with $c\simeq 1.5$ in this particular model). These issues are the
manifestation of the violation of the periodicity discussed in
Sec.~\ref{period}.

\subsection{Model 2}
\label{app:m2}
\begin{figure}
	\begin{center}
		\includegraphics[width=0.99\columnwidth]{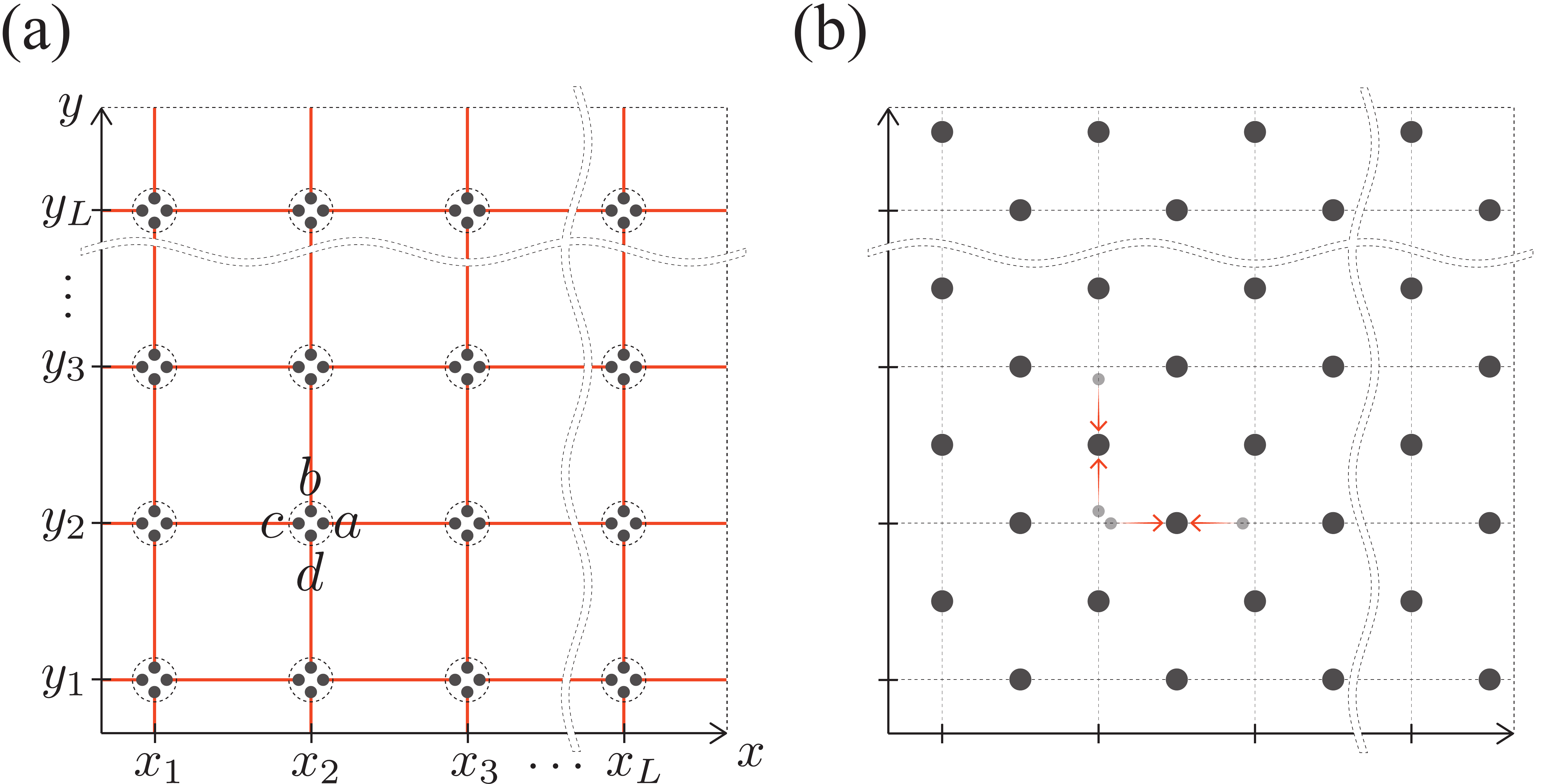}
		\caption{\label{m2} Illustration of the second model (a) and the atomic limit (b).}
	\end{center}
\end{figure}

The second model is the orthogonal stacking of Su-Schrieffer-Heeger chains (Fig.~\ref{m2}). The tight-binding Hamiltonian reads
\begin{align}
H_{\bm k}=&-
\begin{pmatrix}
0&0&e^{ik_x}&0\\
0&0&0&e^{ik_y}\\
e^{-ik_x}&0&0&0\\
0&e^{-ik_y}&0&0\\
\end{pmatrix}.
\label{eq:m2}
\end{align}
This time we have two occupied bands and we place two ions at every lattice site:
\begin{align}
\tilde{q}_{xy,\text{ions}}{'}&=2\tilde{q}_{xy,\text{ions}}=\frac{(2x_1+L-1)^2}{2}.
\end{align}

We analyze this model in the same way as in Appendix~\ref{app2} and find $\Delta\tilde{q}_{xy}^{\text{(a)}}{'}=\tilde{q}_{xy}^{\text{(a)}}{'}-\tilde{q}_{xy,\text{ions}}{'}$. For the atomic limit of this insulator where an electron is localized at both $\bm{x}=\bm{R}+(\frac{1}{2},0)$ and $\bm{R}+(0,\frac{1}{2})$ for every unit cell, we find
\begin{align}
&(\hat{U}_{xy}^{\text{(b)}}{'})_{\bm{R},\bm{R}}=e^{2\pi i[R_x(R_y+1/2)+(R_x+1/2)R_y]/L^2},\label{Mxyd}\\
&\tilde{q}_{xy}^{\text{(b)}}{'}=\frac{1}{2\pi}\sum_{\bm{R}}\text{Im}\ln(\hat{U}_{xy}^{\text{(b)}}{'})_{\bm{R},\bm{R}}=\frac{(2x_1+L-1)(2x_1+L)}{2}.
\end{align}
We list $\Delta\tilde{q}_{xy}^{\text{(b)}}{'}=\tilde{q}_{xy}^{\text{(b)}}{'}-\tilde{q}_{xy,\text{ions}}{'}$ in Table~\ref{tab:m2}. Although we do not expect any quadrupole moment in this model, we found $\Delta\tilde{q}_{xy}=1/2$ mod 1 for the case (b) when $x_1=1$ and $L$ is even.

\begin{table}
\caption{Analytic expression of $\Delta\tilde{q}_{xy}$ for the second model in the limit of large $L$.
\label{tab:m2}}
\begin{tabular}{c|ccc}
\hline\hline
$(x_1,L)$	& \quad$(1,\text{even})$\quad\quad			&	\quad$(1,\text{odd})$\quad\quad			&	\quad$(-\frac{L-1}{2},\text{odd})$\quad\quad			\\\hline
 $\Delta\tilde{q}_{xy}^{\text{(a)}}{'}$	&	$0$	&	$0$	&	$0$\\
 $\Delta\tilde{q}_{xy}^{\text{(b)}}{'}$	&	$\frac{1}{2}$			&	$0$	&	$0$\\
\hline\hline
\end{tabular}
\end{table}

\subsection{Stacking of the two models}
Band insulators considered in Appendices~\ref{app:m1} and \ref{app:m2} possesse a nonzero bulk polarization.
We can fix it by stacking them together.  We show the values of $\Delta\tilde{q}_{xy}$ in the large $L$ limit in Table~\ref{tab:model2D4}.  
As we can see, problems of $\tilde{q}_{xy}$ still persist in the absence of the bulk polarization.

\begin{table}[b]
\caption{The analytic expression of $\Delta\tilde{q}_{xy}$ for the stacked model in the limit of large $L$.
\label{tab:model2D4}}
\begin{tabular}{c|ccc}
\hline\hline
$(x_1,L)$	& $(1,\text{even})$			&	\quad$(1,\text{odd})$\quad\quad			&	\quad$(-\frac{L-1}{2},\text{odd})$\quad\quad			\\\hline
$\Delta\tilde{q}_{xy}^{\text{(a)}}+\Delta\tilde{q}_{xy}^{\text{(a)}}{'}$	&	$\frac{1}{4}+\frac{1}{\pi}\arctan\frac{2}{\pi}$	&	$\frac{3}{4}$		&	$0$\\
$\Delta\tilde{q}_{xy}^{\text{(b)}}+\Delta\tilde{q}_{xy}^{\text{(b)}}{'}$	&	$\frac{1}{4}$			&	$\frac{1}{4}$	&	$\frac{1}{4}$\\
\hline\hline
\end{tabular}
\end{table}

Let us summarize what we learned about $\tilde{q}_{xy}$ through these examples: (i) $\Delta\tilde q_{xy}=\tilde
q_{xy}-\tilde q_{xy,\text{ions}}$ depends on $x_1$ and
the parity of $L$; (ii) it is not always quantized even under the $C_4$ symmetry; and (iii) it takes different values for two states (a) and (b)
in the same phase (i.e., adiabatically connected). 
\bibliography{ref}

\begin{thebibliography}{40}%
\makeatletter
\providecommand \@ifxundefined [1]{%
 \@ifx{#1\undefined}
}%
\providecommand \@ifnum [1]{%
 \ifnum #1\expandafter \@firstoftwo
 \else \expandafter \@secondoftwo
 \fi
}%
\providecommand \@ifx [1]{%
 \ifx #1\expandafter \@firstoftwo
 \else \expandafter \@secondoftwo
 \fi
}%
\providecommand \natexlab [1]{#1}%
\providecommand \enquote  [1]{``#1''}%
\providecommand \bibnamefont  [1]{#1}%
\providecommand \bibfnamefont [1]{#1}%
\providecommand \citenamefont [1]{#1}%
\providecommand \href@noop [0]{\@secondoftwo}%
\providecommand \href [0]{\begingroup \@sanitize@url \@href}%
\providecommand \@href[1]{\@@startlink{#1}\@@href}%
\providecommand \@@href[1]{\endgroup#1\@@endlink}%
\providecommand \@sanitize@url [0]{\catcode `\\12\catcode `\$12\catcode
  `\&12\catcode `\#12\catcode `\^12\catcode `\_12\catcode `\%12\relax}%
\providecommand \@@startlink[1]{}%
\providecommand \@@endlink[0]{}%
\providecommand \url  [0]{\begingroup\@sanitize@url \@url }%
\providecommand \@url [1]{\endgroup\@href {#1}{\urlprefix }}%
\providecommand \urlprefix  [0]{URL }%
\providecommand \Eprint [0]{\href }%
\providecommand \doibase [0]{http://dx.doi.org/}%
\providecommand \selectlanguage [0]{\@gobble}%
\providecommand \bibinfo  [0]{\@secondoftwo}%
\providecommand \bibfield  [0]{\@secondoftwo}%
\providecommand \translation [1]{[#1]}%
\providecommand \BibitemOpen [0]{}%
\providecommand \bibitemStop [0]{}%
\providecommand \bibitemNoStop [0]{.\EOS\space}%
\providecommand \EOS [0]{\spacefactor3000\relax}%
\providecommand \BibitemShut  [1]{\csname bibitem#1\endcsname}%
\let\auto@bib@innerbib\@empty
\bibitem [{\citenamefont {Parameswaran}\ and\ \citenamefont
  {Wan}(2017)}]{parameswaran2017}%
  \BibitemOpen
  \bibfield  {author} {\bibinfo {author} {\bibfnamefont {S.~A.}\ \bibnamefont
  {Parameswaran}}\ and\ \bibinfo {author} {\bibfnamefont {Y.}~\bibnamefont
  {Wan}},\ }\href {\doibase 10.1103/Physics.10.132} {\bibfield  {journal}
  {\bibinfo  {journal} {Physics}\ }\textbf {\bibinfo {volume} {10}},\ \bibinfo
  {pages} {132} (\bibinfo {year} {2017})}\BibitemShut {NoStop}%
\bibitem [{\citenamefont {Schindler}\ \emph {et~al.}(2018)\citenamefont
  {Schindler}, \citenamefont {Cook}, \citenamefont {Vergniory}, \citenamefont
  {Wang}, \citenamefont {Parkin}, \citenamefont {Bernevig},\ and\ \citenamefont
  {Neupert}}]{schindler2018}%
  \BibitemOpen
  \bibfield  {author} {\bibinfo {author} {\bibfnamefont {F.}~\bibnamefont
  {Schindler}}, \bibinfo {author} {\bibfnamefont {A.~M.}\ \bibnamefont {Cook}},
  \bibinfo {author} {\bibfnamefont {M.~G.}\ \bibnamefont {Vergniory}}, \bibinfo
  {author} {\bibfnamefont {Z.}~\bibnamefont {Wang}}, \bibinfo {author}
  {\bibfnamefont {S.~S.~P.}\ \bibnamefont {Parkin}}, \bibinfo {author}
  {\bibfnamefont {B.~A.}\ \bibnamefont {Bernevig}}, \ and\ \bibinfo {author}
  {\bibfnamefont {T.}~\bibnamefont {Neupert}},\ }\href
  {http://advances.sciencemag.org/content/4/6/eaat0346} {\bibfield  {journal}
  {\bibinfo  {journal} {Sci. Adv.}\ }\textbf {\bibinfo {volume} {4}},\ \bibinfo
  {pages} {eaat0346} (\bibinfo {year} {2018})}\BibitemShut {NoStop}%
\bibitem [{\citenamefont {Peng}\ \emph {et~al.}(2017)\citenamefont {Peng},
  \citenamefont {Bao},\ and\ \citenamefont {von Oppen}}]{peng2017}%
  \BibitemOpen
  \bibfield  {author} {\bibinfo {author} {\bibfnamefont {Y.}~\bibnamefont
  {Peng}}, \bibinfo {author} {\bibfnamefont {Y.}~\bibnamefont {Bao}}, \ and\
  \bibinfo {author} {\bibfnamefont {F.}~\bibnamefont {von Oppen}},\ }\href
  {\doibase 10.1103/PhysRevB.95.235143} {\bibfield  {journal} {\bibinfo
  {journal} {Phys. Rev. B}\ }\textbf {\bibinfo {volume} {95}},\ \bibinfo
  {pages} {235143} (\bibinfo {year} {2017})}\BibitemShut {NoStop}%
\bibitem [{\citenamefont {Langbehn}\ \emph {et~al.}(2017)\citenamefont
  {Langbehn}, \citenamefont {Peng}, \citenamefont {Trifunovic}, \citenamefont
  {von Oppen},\ and\ \citenamefont {Brouwer}}]{langbehn2017}%
  \BibitemOpen
  \bibfield  {author} {\bibinfo {author} {\bibfnamefont {J.}~\bibnamefont
  {Langbehn}}, \bibinfo {author} {\bibfnamefont {Y.}~\bibnamefont {Peng}},
  \bibinfo {author} {\bibfnamefont {L.}~\bibnamefont {Trifunovic}}, \bibinfo
  {author} {\bibfnamefont {F.}~\bibnamefont {von Oppen}}, \ and\ \bibinfo
  {author} {\bibfnamefont {P.~W.}\ \bibnamefont {Brouwer}},\ }\href {\doibase
  10.1103/PhysRevLett.119.246401} {\bibfield  {journal} {\bibinfo  {journal}
  {Phys. Rev. Lett.}\ }\textbf {\bibinfo {volume} {119}},\ \bibinfo {pages}
  {246401} (\bibinfo {year} {2017})}\BibitemShut {NoStop}%
\bibitem [{\citenamefont {Song}\ \emph {et~al.}(2017)\citenamefont {Song},
  \citenamefont {Fang},\ and\ \citenamefont {Fang}}]{song2017}%
  \BibitemOpen
  \bibfield  {author} {\bibinfo {author} {\bibfnamefont {Z.}~\bibnamefont
  {Song}}, \bibinfo {author} {\bibfnamefont {Z.}~\bibnamefont {Fang}}, \ and\
  \bibinfo {author} {\bibfnamefont {C.}~\bibnamefont {Fang}},\ }\href {\doibase
  10.1103/PhysRevLett.119.246402} {\bibfield  {journal} {\bibinfo  {journal}
  {Phys. Rev. Lett.}\ }\textbf {\bibinfo {volume} {119}},\ \bibinfo {pages}
  {246402} (\bibinfo {year} {2017})}\BibitemShut {NoStop}%
\bibitem [{\citenamefont {Benalcazar}\ \emph
  {et~al.}(2017{\natexlab{a}})\citenamefont {Benalcazar}, \citenamefont
  {Bernevig},\ and\ \citenamefont {Hughes}}]{Benalcazar61}%
  \BibitemOpen
  \bibfield  {author} {\bibinfo {author} {\bibfnamefont {W.~A.}\ \bibnamefont
  {Benalcazar}}, \bibinfo {author} {\bibfnamefont {B.~A.}\ \bibnamefont
  {Bernevig}}, \ and\ \bibinfo {author} {\bibfnamefont {T.~L.}\ \bibnamefont
  {Hughes}},\ }\href {\doibase 10.1126/science.aah6442} {\bibfield  {journal}
  {\bibinfo  {journal} {Science}\ }\textbf {\bibinfo {volume} {357}},\ \bibinfo
  {pages} {61} (\bibinfo {year} {2017}{\natexlab{a}})}\BibitemShut {NoStop}%
\bibitem [{\citenamefont {Benalcazar}\ \emph
  {et~al.}(2017{\natexlab{b}})\citenamefont {Benalcazar}, \citenamefont
  {Bernevig},\ and\ \citenamefont {Hughes}}]{PhysRevB.96.245115}%
  \BibitemOpen
  \bibfield  {author} {\bibinfo {author} {\bibfnamefont {W.~A.}\ \bibnamefont
  {Benalcazar}}, \bibinfo {author} {\bibfnamefont {B.~A.}\ \bibnamefont
  {Bernevig}}, \ and\ \bibinfo {author} {\bibfnamefont {T.~L.}\ \bibnamefont
  {Hughes}},\ }\href {\doibase 10.1103/PhysRevB.96.245115} {\bibfield
  {journal} {\bibinfo  {journal} {Phys. Rev. B}\ }\textbf {\bibinfo {volume}
  {96}},\ \bibinfo {pages} {245115} (\bibinfo {year}
  {2017}{\natexlab{b}})}\BibitemShut {NoStop}%
\bibitem [{\citenamefont {Fang}\ and\ \citenamefont {Fu}(2017)}]{fang2018}%
  \BibitemOpen
  \bibfield  {author} {\bibinfo {author} {\bibfnamefont {C.}~\bibnamefont
  {Fang}}\ and\ \bibinfo {author} {\bibfnamefont {L.}~\bibnamefont {Fu}},\
  }\href@noop {} {\bibfield  {journal} {\bibinfo  {journal} {arXiv:1709.01929}\
  } (\bibinfo {year} {2017})}\BibitemShut {NoStop}%
\bibitem [{\citenamefont {Ezawa}(2018)}]{ezawa2018}%
  \BibitemOpen
  \bibfield  {author} {\bibinfo {author} {\bibfnamefont {M.}~\bibnamefont
  {Ezawa}},\ }\href {\doibase 10.1103/PhysRevLett.120.026801} {\bibfield
  {journal} {\bibinfo  {journal} {Phys. Rev. Lett.}\ }\textbf {\bibinfo
  {volume} {120}},\ \bibinfo {pages} {026801} (\bibinfo {year}
  {2018})}\BibitemShut {NoStop}%
\bibitem [{\citenamefont {Shapourian}\ \emph {et~al.}(2018)\citenamefont
  {Shapourian}, \citenamefont {Wang},\ and\ \citenamefont
  {Ryu}}]{shapourian2017}%
  \BibitemOpen
  \bibfield  {author} {\bibinfo {author} {\bibfnamefont {H.}~\bibnamefont
  {Shapourian}}, \bibinfo {author} {\bibfnamefont {Y.}~\bibnamefont {Wang}}, \
  and\ \bibinfo {author} {\bibfnamefont {S.}~\bibnamefont {Ryu}},\ }\href
  {\doibase 10.1103/PhysRevB.97.094508} {\bibfield  {journal} {\bibinfo
  {journal} {Phys. Rev. B}\ }\textbf {\bibinfo {volume} {97}},\ \bibinfo
  {pages} {094508} (\bibinfo {year} {2018})}\BibitemShut {NoStop}%
\bibitem [{\citenamefont {Zhu}(2018)}]{zhu2018}%
  \BibitemOpen
  \bibfield  {author} {\bibinfo {author} {\bibfnamefont {X.}~\bibnamefont
  {Zhu}},\ }\href {\doibase 10.1103/PhysRevB.97.205134} {\bibfield  {journal}
  {\bibinfo  {journal} {Phys. Rev. B}\ }\textbf {\bibinfo {volume} {97}},\
  \bibinfo {pages} {205134} (\bibinfo {year} {2018})}\BibitemShut {NoStop}%
\bibitem [{\citenamefont {Yan}\ \emph {et~al.}(2018)\citenamefont {Yan},
  \citenamefont {Song},\ and\ \citenamefont {Wang}}]{yan2018}%
  \BibitemOpen
  \bibfield  {author} {\bibinfo {author} {\bibfnamefont {Z.}~\bibnamefont
  {Yan}}, \bibinfo {author} {\bibfnamefont {F.}~\bibnamefont {Song}}, \ and\
  \bibinfo {author} {\bibfnamefont {Z.}~\bibnamefont {Wang}},\ }\href {\doibase
  10.1103/PhysRevLett.121.096803} {\bibfield  {journal} {\bibinfo  {journal}
  {Phys. Rev. Lett.}\ }\textbf {\bibinfo {volume} {121}},\ \bibinfo {pages}
  {096803} (\bibinfo {year} {2018})}\BibitemShut {NoStop}%
\bibitem [{\citenamefont {{Wang}}\ \emph
  {et~al.}(2018{\natexlab{a}})\citenamefont {{Wang}}, \citenamefont {{Lin}},\
  and\ \citenamefont {{Hughes}}}]{wang2018}%
  \BibitemOpen
  \bibfield  {author} {\bibinfo {author} {\bibfnamefont {Y.}~\bibnamefont
  {{Wang}}}, \bibinfo {author} {\bibfnamefont {M.}~\bibnamefont {{Lin}}}, \
  and\ \bibinfo {author} {\bibfnamefont {T.~L.}\ \bibnamefont {{Hughes}}},\
  }\href@noop {} {\bibfield  {journal} {\bibinfo  {journal} {ArXiv e-prints}\ }
  (\bibinfo {year} {2018}{\natexlab{a}})},\ \Eprint
  {http://arxiv.org/abs/1804.01531} {arXiv:1804.01531 [cond-mat.supr-con]}
  \BibitemShut {NoStop}%
\bibitem [{\citenamefont {{Wang}}\ \emph
  {et~al.}(2018{\natexlab{b}})\citenamefont {{Wang}}, \citenamefont {{Liu}},
  \citenamefont {{Lu}},\ and\ \citenamefont {{Zhang}}}]{wang2018b}%
  \BibitemOpen
  \bibfield  {author} {\bibinfo {author} {\bibfnamefont {Q.}~\bibnamefont
  {{Wang}}}, \bibinfo {author} {\bibfnamefont {C.-C.}\ \bibnamefont {{Liu}}},
  \bibinfo {author} {\bibfnamefont {Y.-M.}\ \bibnamefont {{Lu}}}, \ and\
  \bibinfo {author} {\bibfnamefont {F.}~\bibnamefont {{Zhang}}},\ }\href@noop
  {} {\bibfield  {journal} {\bibinfo  {journal} {ArXiv e-prints}\ } (\bibinfo
  {year} {2018}{\natexlab{b}})},\ \Eprint {http://arxiv.org/abs/1804.04711}
  {arXiv:1804.04711 [cond-mat.mes-hall]} \BibitemShut {NoStop}%
\bibitem [{\citenamefont {Khalaf}\ \emph {et~al.}(2018)\citenamefont {Khalaf},
  \citenamefont {Po}, \citenamefont {Vishwanath},\ and\ \citenamefont
  {Watanabe}}]{khalaf2018}%
  \BibitemOpen
  \bibfield  {author} {\bibinfo {author} {\bibfnamefont {E.}~\bibnamefont
  {Khalaf}}, \bibinfo {author} {\bibfnamefont {H.~C.}\ \bibnamefont {Po}},
  \bibinfo {author} {\bibfnamefont {A.}~\bibnamefont {Vishwanath}}, \ and\
  \bibinfo {author} {\bibfnamefont {H.}~\bibnamefont {Watanabe}},\ }\href
  {\doibase 10.1103/PhysRevX.8.031070} {\bibfield  {journal} {\bibinfo
  {journal} {Phys. Rev. X}\ }\textbf {\bibinfo {volume} {8}},\ \bibinfo {pages}
  {031070} (\bibinfo {year} {2018})}\BibitemShut {NoStop}%
\bibitem [{\citenamefont {Khalaf}(2018)}]{khalaf2018b}%
  \BibitemOpen
  \bibfield  {author} {\bibinfo {author} {\bibfnamefont {E.}~\bibnamefont
  {Khalaf}},\ }\href {\doibase 10.1103/PhysRevB.97.205136} {\bibfield
  {journal} {\bibinfo  {journal} {Phys. Rev. B}\ }\textbf {\bibinfo {volume}
  {97}},\ \bibinfo {pages} {205136} (\bibinfo {year} {2018})}\BibitemShut
  {NoStop}%
\bibitem [{\citenamefont {Trifunovic}\ and\ \citenamefont
  {Brouwer}(2019)}]{trifunovic2019}%
  \BibitemOpen
  \bibfield  {author} {\bibinfo {author} {\bibfnamefont {L.}~\bibnamefont
  {Trifunovic}}\ and\ \bibinfo {author} {\bibfnamefont {P.~W.}\ \bibnamefont
  {Brouwer}},\ }\href {\doibase 10.1103/PhysRevX.9.011012} {\bibfield
  {journal} {\bibinfo  {journal} {Phys. Rev. X}\ }\textbf {\bibinfo {volume}
  {9}},\ \bibinfo {pages} {011012} (\bibinfo {year} {2019})}\BibitemShut
  {NoStop}%
\bibitem [{\citenamefont {{Okuma}}\ \emph {et~al.}(2018)\citenamefont
  {{Okuma}}, \citenamefont {{Sato}},\ and\ \citenamefont
  {{Shiozaki}}}]{nobuyuki2018}%
  \BibitemOpen
  \bibfield  {author} {\bibinfo {author} {\bibfnamefont {N.}~\bibnamefont
  {{Okuma}}}, \bibinfo {author} {\bibfnamefont {M.}~\bibnamefont {{Sato}}}, \
  and\ \bibinfo {author} {\bibfnamefont {K.}~\bibnamefont {{Shiozaki}}},\
  }\href@noop {} {\bibfield  {journal} {\bibinfo  {journal} {arXiv e-prints}\
  ,\ \bibinfo {eid} {arXiv:1810.12601}} (\bibinfo {year} {2018})},\ \Eprint
  {http://arxiv.org/abs/1810.12601} {arXiv:1810.12601 [cond-mat.mes-hall]}
  \BibitemShut {NoStop}%
\bibitem [{\citenamefont {Resta}\ and\ \citenamefont
  {Vanderbilt}(2007)}]{Resta2007}%
  \BibitemOpen
  \bibfield  {author} {\bibinfo {author} {\bibfnamefont {R.}~\bibnamefont
  {Resta}}\ and\ \bibinfo {author} {\bibfnamefont {D.}~\bibnamefont
  {Vanderbilt}},\ }\enquote {\bibinfo {title} {Theory of polarization: A modern
  approach},}\ in\ \href {\doibase 10.1007/978-3-540-34591-6_2} {\emph
  {\bibinfo {booktitle} {Physics of Ferroelectrics: A Modern Perspective}}}\
  (\bibinfo  {publisher} {Springer Berlin Heidelberg},\ \bibinfo {address}
  {Berlin, Heidelberg},\ \bibinfo {year} {2007})\ pp.\ \bibinfo {pages}
  {31--68}\BibitemShut {NoStop}%
\bibitem [{\citenamefont {Lines}\ and\ \citenamefont {Glass}(2001)}]{Lines}%
  \BibitemOpen
  \bibfield  {author} {\bibinfo {author} {\bibfnamefont {M.~E.}\ \bibnamefont
  {Lines}}\ and\ \bibinfo {author} {\bibfnamefont {A.~M.}\ \bibnamefont
  {Glass}},\ }\href@noop {} {\emph {\bibinfo {title} {Principles and
  Applications of Ferroelectrics and Related Materials}}}\ (\bibinfo
  {publisher} {Oxford University Press},\ \bibinfo {year} {2001})\BibitemShut
  {NoStop}%
\bibitem [{\citenamefont {Kittel}(2004)}]{kittel}%
  \BibitemOpen
  \bibfield  {author} {\bibinfo {author} {\bibfnamefont {C.}~\bibnamefont
  {Kittel}},\ }\href@noop {} {\emph {\bibinfo {title} {Introduction to Solid
  State Physics}}}\ (\bibinfo  {publisher} {Wiley},\ \bibinfo {year}
  {2004})\BibitemShut {NoStop}%
\bibitem [{\citenamefont {Martin}(1974)}]{Martin}%
  \BibitemOpen
  \bibfield  {author} {\bibinfo {author} {\bibfnamefont {R.~M.}\ \bibnamefont
  {Martin}},\ }\href {\doibase 10.1103/PhysRevB.9.1998} {\bibfield  {journal}
  {\bibinfo  {journal} {Phys. Rev. B}\ }\textbf {\bibinfo {volume} {9}},\
  \bibinfo {pages} {1998} (\bibinfo {year} {1974})}\BibitemShut {NoStop}%
\bibitem [{\citenamefont {Aligia}\ and\ \citenamefont
  {Ortiz}(1999)}]{PhysRevLett.82.2560}%
  \BibitemOpen
  \bibfield  {author} {\bibinfo {author} {\bibfnamefont {A.~A.}\ \bibnamefont
  {Aligia}}\ and\ \bibinfo {author} {\bibfnamefont {G.}~\bibnamefont {Ortiz}},\
  }\href {\doibase 10.1103/PhysRevLett.82.2560} {\bibfield  {journal} {\bibinfo
   {journal} {Phys. Rev. Lett.}\ }\textbf {\bibinfo {volume} {82}},\ \bibinfo
  {pages} {2560} (\bibinfo {year} {1999})}\BibitemShut {NoStop}%
\bibitem [{\citenamefont {Vanderbilt}\ and\ \citenamefont
  {King-Smith}(1993)}]{PhysRevB.48.4442}%
  \BibitemOpen
  \bibfield  {author} {\bibinfo {author} {\bibfnamefont {D.}~\bibnamefont
  {Vanderbilt}}\ and\ \bibinfo {author} {\bibfnamefont {R.~D.}\ \bibnamefont
  {King-Smith}},\ }\href {\doibase 10.1103/PhysRevB.48.4442} {\bibfield
  {journal} {\bibinfo  {journal} {Phys. Rev. B}\ }\textbf {\bibinfo {volume}
  {48}},\ \bibinfo {pages} {4442} (\bibinfo {year} {1993})}\BibitemShut
  {NoStop}%
\bibitem [{\citenamefont {King-Smith}\ and\ \citenamefont
  {Vanderbilt}(1993)}]{PhysRevB.47.1651}%
  \BibitemOpen
  \bibfield  {author} {\bibinfo {author} {\bibfnamefont {R.~D.}\ \bibnamefont
  {King-Smith}}\ and\ \bibinfo {author} {\bibfnamefont {D.}~\bibnamefont
  {Vanderbilt}},\ }\href {\doibase 10.1103/PhysRevB.47.1651} {\bibfield
  {journal} {\bibinfo  {journal} {Phys. Rev. B}\ }\textbf {\bibinfo {volume}
  {47}},\ \bibinfo {pages} {1651} (\bibinfo {year} {1993})}\BibitemShut
  {NoStop}%
\bibitem [{\citenamefont {Ortiz}\ and\ \citenamefont
  {Martin}(1994)}]{PhysRevB.49.14202}%
  \BibitemOpen
  \bibfield  {author} {\bibinfo {author} {\bibfnamefont {G.}~\bibnamefont
  {Ortiz}}\ and\ \bibinfo {author} {\bibfnamefont {R.~M.}\ \bibnamefont
  {Martin}},\ }\href {\doibase 10.1103/PhysRevB.49.14202} {\bibfield  {journal}
  {\bibinfo  {journal} {Phys. Rev. B}\ }\textbf {\bibinfo {volume} {49}},\
  \bibinfo {pages} {14202} (\bibinfo {year} {1994})}\BibitemShut {NoStop}%
\bibitem [{\citenamefont {Souza}\ \emph {et~al.}(2000)\citenamefont {Souza},
  \citenamefont {Wilkens},\ and\ \citenamefont {Martin}}]{Souza}%
  \BibitemOpen
  \bibfield  {author} {\bibinfo {author} {\bibfnamefont {I.}~\bibnamefont
  {Souza}}, \bibinfo {author} {\bibfnamefont {T.}~\bibnamefont {Wilkens}}, \
  and\ \bibinfo {author} {\bibfnamefont {R.~M.}\ \bibnamefont {Martin}},\
  }\href {\doibase 10.1103/PhysRevB.62.1666} {\bibfield  {journal} {\bibinfo
  {journal} {Phys. Rev. B}\ }\textbf {\bibinfo {volume} {62}},\ \bibinfo
  {pages} {1666} (\bibinfo {year} {2000})}\BibitemShut {NoStop}%
\bibitem [{\citenamefont {Watanabe}\ and\ \citenamefont
  {Oshikawa}(2018)}]{PhysRevX.8.021065}%
  \BibitemOpen
  \bibfield  {author} {\bibinfo {author} {\bibfnamefont {H.}~\bibnamefont
  {Watanabe}}\ and\ \bibinfo {author} {\bibfnamefont {M.}~\bibnamefont
  {Oshikawa}},\ }\href {\doibase 10.1103/PhysRevX.8.021065} {\bibfield
  {journal} {\bibinfo  {journal} {Phys. Rev. X}\ }\textbf {\bibinfo {volume}
  {8}},\ \bibinfo {pages} {021065} (\bibinfo {year} {2018})}\BibitemShut
  {NoStop}%
\bibitem [{\citenamefont {Resta}(1998)}]{Resta}%
  \BibitemOpen
  \bibfield  {author} {\bibinfo {author} {\bibfnamefont {R.}~\bibnamefont
  {Resta}},\ }\href {\doibase 10.1103/PhysRevLett.80.1800} {\bibfield
  {journal} {\bibinfo  {journal} {Phys. Rev. Lett.}\ }\textbf {\bibinfo
  {volume} {80}},\ \bibinfo {pages} {1800} (\bibinfo {year}
  {1998})}\BibitemShut {NoStop}%
\bibitem [{\citenamefont {Trifunovic}\ \emph {et~al.}(2019)\citenamefont
  {Trifunovic}, \citenamefont {Ono},\ and\ \citenamefont
  {Watanabe}}]{trifunovic2019b}%
  \BibitemOpen
  \bibfield  {author} {\bibinfo {author} {\bibfnamefont {L.}~\bibnamefont
  {Trifunovic}}, \bibinfo {author} {\bibfnamefont {S.}~\bibnamefont {Ono}}, \
  and\ \bibinfo {author} {\bibfnamefont {H.}~\bibnamefont {Watanabe}},\ }\href
  {\doibase 10.1103/PhysRevB.100.054408} {\bibfield  {journal} {\bibinfo
  {journal} {Phys. Rev. B}\ }\textbf {\bibinfo {volume} {100}},\ \bibinfo
  {pages} {054408} (\bibinfo {year} {2019})}\BibitemShut {NoStop}%
\bibitem [{\citenamefont {Wheeler}\ \emph {et~al.}(2018)\citenamefont
  {Wheeler}, \citenamefont {Wagner},\ and\ \citenamefont {Hughes}}]{Taylor}%
  \BibitemOpen
  \bibfield  {author} {\bibinfo {author} {\bibfnamefont {W.~A.}\ \bibnamefont
  {Wheeler}}, \bibinfo {author} {\bibfnamefont {L.~K.}\ \bibnamefont {Wagner}},
  \ and\ \bibinfo {author} {\bibfnamefont {T.~L.}\ \bibnamefont {Hughes}},\
  }\href@noop {} {} (\bibinfo {year} {2018}),\ \Eprint
  {http://arxiv.org/abs/arXiv:1812.06990} {arXiv:1812.06990} \BibitemShut
  {NoStop}%
\bibitem [{\citenamefont {Kang}\ \emph {et~al.}(2018)\citenamefont {Kang},
  \citenamefont {Shiozaki},\ and\ \citenamefont {Cho}}]{Ken}%
  \BibitemOpen
  \bibfield  {author} {\bibinfo {author} {\bibfnamefont {B.}~\bibnamefont
  {Kang}}, \bibinfo {author} {\bibfnamefont {K.}~\bibnamefont {Shiozaki}}, \
  and\ \bibinfo {author} {\bibfnamefont {G.~Y.}\ \bibnamefont {Cho}},\
  }\href@noop {} {} (\bibinfo {year} {2018}),\ \Eprint
  {http://arxiv.org/abs/arXiv:1812.06999} {arXiv:1812.06999} \BibitemShut
  {NoStop}%
\bibitem [{Note1()}]{Note1}%
  \BibitemOpen
  \bibinfo {note} {We explain why this perturbation theory fails in
  Appendix~\ref {app1}.}\BibitemShut {Stop}%
\bibitem [{\citenamefont {{Agarwala}}\ \emph {et~al.}(2019)\citenamefont
  {{Agarwala}}, \citenamefont {{Juricic}},\ and\ \citenamefont
  {{Roy}}}]{agarwala2019}%
  \BibitemOpen
  \bibfield  {author} {\bibinfo {author} {\bibfnamefont {A.}~\bibnamefont
  {{Agarwala}}}, \bibinfo {author} {\bibfnamefont {V.}~\bibnamefont
  {{Juricic}}}, \ and\ \bibinfo {author} {\bibfnamefont {B.}~\bibnamefont
  {{Roy}}},\ }\href@noop {} {\bibfield  {journal} {\bibinfo  {journal} {arXiv
  e-prints}\ ,\ \bibinfo {eid} {arXiv:1902.00507}} (\bibinfo {year} {2019})},\
  \Eprint {http://arxiv.org/abs/1902.00507} {arXiv:1902.00507
  [cond-mat.mes-hall]} \BibitemShut {NoStop}%
\bibitem [{\citenamefont {{Lin}}\ \emph {et~al.}(2019)\citenamefont {{Lin}},
  \citenamefont {{Wang}}, \citenamefont {{Lu}},\ and\ \citenamefont
  {{Jiang}}}]{lin2019}%
  \BibitemOpen
  \bibfield  {author} {\bibinfo {author} {\bibfnamefont {Z.-K.}\ \bibnamefont
  {{Lin}}}, \bibinfo {author} {\bibfnamefont {H.-X.}\ \bibnamefont {{Wang}}},
  \bibinfo {author} {\bibfnamefont {M.-H.}\ \bibnamefont {{Lu}}}, \ and\
  \bibinfo {author} {\bibfnamefont {J.-H.}\ \bibnamefont {{Jiang}}},\
  }\href@noop {} {\bibfield  {journal} {\bibinfo  {journal} {arXiv e-prints}\
  ,\ \bibinfo {eid} {arXiv:1903.05997}} (\bibinfo {year} {2019})},\ \Eprint
  {http://arxiv.org/abs/1903.05997} {arXiv:1903.05997 [cond-mat.str-el]}
  \BibitemShut {NoStop}%
\bibitem [{\citenamefont {Benalcazar}\ \emph {et~al.}(2019)\citenamefont
  {Benalcazar}, \citenamefont {Li},\ and\ \citenamefont {Hughes}}]{1809.02142}%
  \BibitemOpen
  \bibfield  {author} {\bibinfo {author} {\bibfnamefont {W.~A.}\ \bibnamefont
  {Benalcazar}}, \bibinfo {author} {\bibfnamefont {T.}~\bibnamefont {Li}}, \
  and\ \bibinfo {author} {\bibfnamefont {T.~L.}\ \bibnamefont {Hughes}},\
  }\href {\doibase 10.1103/PhysRevB.99.245151} {\bibfield  {journal} {\bibinfo
  {journal} {Phys. Rev. B}\ }\textbf {\bibinfo {volume} {99}},\ \bibinfo
  {pages} {245151} (\bibinfo {year} {2019})}\BibitemShut {NoStop}%
\bibitem [{Note2()}]{Note2}%
  \BibitemOpen
  \bibinfo {note} {See also Appendix of Ref.~\protect \rev@citealpnum {Ken}
  where the dependence of $x_1$ is discussed using a tight-binding
  model.}\BibitemShut {Stop}%
\bibitem [{\citenamefont {Marzari}\ and\ \citenamefont
  {Vanderbilt}(1997)}]{Marzari}%
  \BibitemOpen
  \bibfield  {author} {\bibinfo {author} {\bibfnamefont {N.}~\bibnamefont
  {Marzari}}\ and\ \bibinfo {author} {\bibfnamefont {D.}~\bibnamefont
  {Vanderbilt}},\ }\href {\doibase 10.1103/PhysRevB.56.12847} {\bibfield
  {journal} {\bibinfo  {journal} {Phys. Rev. B}\ }\textbf {\bibinfo {volume}
  {56}},\ \bibinfo {pages} {12847} (\bibinfo {year} {1997})}\BibitemShut
  {NoStop}%
\bibitem [{\citenamefont {Resta}\ and\ \citenamefont {Sorella}(1999)}]{Resta2}%
  \BibitemOpen
  \bibfield  {author} {\bibinfo {author} {\bibfnamefont {R.}~\bibnamefont
  {Resta}}\ and\ \bibinfo {author} {\bibfnamefont {S.}~\bibnamefont
  {Sorella}},\ }\href {\doibase 10.1103/PhysRevLett.82.370} {\bibfield
  {journal} {\bibinfo  {journal} {Phys. Rev. Lett.}\ }\textbf {\bibinfo
  {volume} {82}},\ \bibinfo {pages} {370} (\bibinfo {year} {1999})}\BibitemShut
  {NoStop}%
\bibitem [{Note3()}]{Note3}%
  \BibitemOpen
  \bibinfo {note} {Even when $d=1$, Eq.~\protect \textup {\hbox {\mathsurround
  \z@ \protect \normalfont (\ignorespaces \ref {wrong}\unskip \@@italiccorr )}}
  is still violated because the exponent of Eq.~\protect \textup {\hbox
  {\mathsurround \z@ \protect \normalfont (\ignorespaces \ref
  {correctscale}\unskip \@@italiccorr )}} is $O(L_x^{-1})$, not
  $O(L_x^{-2})$.}\BibitemShut {Stop}%
\end{thebibliography}%

\end{document}